\documentclass[aps,twocolumn,epsfig,graphics,showpacs,floatfix]{revtex4}

\usepackage{amsmath,amsfonts,amssymb,graphics,graphicx,epsfig,color,times,bbm}

\begin{document}

\bibliographystyle{apsrev}

\newtheorem{theorem}{Theorem}
\newtheorem{proposition}{Proposition}
\newtheorem{lemma}{Lemma}

\newcommand{\proofend}{\hfill\fbox\\\medskip }
\newcommand{\tr}{{\text{tr}}}
\newcommand{\id}{{\mathbbm{1}}}
\newcommand{\RR}{{\mathbbm{R}}}
\newcommand{\CC}{{\mathbbm{C}}}
\newcommand{\NN}{{\mathbbm{N}}}

\newcommand{\proof}[1]{{\bf Proof.} #1 $\proofend$}

\title{Optimizing Linear Optics Quantum Gates}

\author{J.\ Eisert}

\affiliation{1 QOLS, Blackett Laboratory, Imperial College London, London, SW7 2BW, 
UK\\
2 Institute for Mathematical Sciences, Imperial College London, London, SW7 2BW, 
UK\\
3 Institut f{\"u}r Physik, Universit{\"a}t Potsdam, 14469 
Potsdam, Germany}

\begin{abstract}
In this paper, the problem of finding 
optimal success probabilities of static 
linear optics quantum gates 
is linked to the theory of convex optimization. 
It is shown that by exploiting this link,
upper bounds for the success probability of 
networks realizing single-mode gates can be derived, 
which hold in generality for linear optical networks followed by postselection,
i.e., for networks of arbitrary size, 
any number of auxiliary modes, and arbitrary photon numbers.
As a corollary, the previously formulated conjecture
is proven that the optimal success
probability of a postselected non-linear sign shift 
without feed-forward is $1/4$, a gate playing
the central role in the scheme 
of Knill-Laflamme-Milburn for quantum computation with
linear optics. The concept of Lagrange duality 
is shown to be applicable to
provide rigorous proofs for such bounds 
for elementary gates, 
although the original problem is a difficult
non-convex problem in infinitely many objective variables. 
The versatility of this approach to identify 
other optimal linear optical schemes 
is demonstrated.

\end{abstract}

\maketitle


Optical implementations of quantum information processing
devices offer many advantages over implementations employing other 
physical systems. Photons are relatively
prone to decoherence, and precise state control is possible with the help 
of linear optical elements \cite{Mike}. Moreover, although 
 the required non-linearities to do 
universal quantum computation are presently not available at the single-photon level, 
they can be effectively realized by means of 
measurements.
This comes at the price of
the scheme becoming probabilistic. It was one of the key insights
in the field,
proposed by Knill, Laflamme, and Milburn \cite{KLM}, that
quantum computation can be achieved
in a near-deterministic way using only single photon sources, linear optical 
elements and photon counters \cite{KLM,Knill,Pittman}. For this to be possible
a significant overhead in resources is required \cite{KLM,Cluster}. 
At the basis of the construction of the original scheme, however, is 
a gate that is 
implemented with some probability of success, the non-linear sign shift 
gate \cite{KLM,Knill,Kojima}. The best known success probability of this gate using 
static linear optics 
followed by postselection is one quarter; this 
can then be  uplifted to close to unity using teleportation steps.

One of the central questions seems therefore: how well can the elementary 
gates be performed with static linear optics networks?
In particular, what are the upper bounds for success probabilities 
of energy-preserving gates of single-modes? 
This seems a key question for two reasons: 
on the one hand, the success
probability at the level of elementary gates is a quantity that 
determines the necessary and notably large 
overhead to achieve near-deterministic scalable quantum computation \cite{KLM,ScalingRalph}. 
On the other hand, for small-scale applications such as 
quantum repeaters for the long-range distribution of entanglement, high fidelity of 
the quantum gates may often be the demanding requirement 
of salient interest. 
The achievable rates in entanglement distillation, 
say, may be of secondary importance compared to the 
very functioning of the scheme. 
In such contexts, one should be expected to be well advised
to abandon some of the feed-forward using 
quantum memories or delay lines, but rather postselect the outcomes. 

The best known scheme to realize the non-linear sign shift gate 
with linear optics without  feed-forward
succeeds with a probability of a quarter. 
Later Knill showed  that  the success probability 
can at most reach one half \cite{Knill}. This was an 
important step: it was not clear, yet, whether this bound was
indeed tight. Aiming at tightening this bound, Scheel und 
L{\"u}tkenhaus made a further significant
step, emphasising that a linear optics network realizing a 
quantum gate can be thought of as one which is  
linked once to the input mode by a single
beam splitter \cite{ScheelLuetke}, based on a result by
Reck, Zeilinger and co-workers \cite{Reck,Ulf,Knight}.
It was conjectured, based on a numerical analysis 
in a restricted setting,
that the maximal success probability of this gate
could be one quarter. 

It is the aim of this paper to link the question of success probabilities
to the theory of convex optimization 
\cite{ConvexButNotSemidef,ConvexInQuantumInfo,ConvexInQuantumInfoOther,Semi}. 
It turns out that   convex optimization 
provides powerful analytical methods to 
prove the validity of bounds to 
optimal success probabilities, without having to resort 
to restrictions of generality. 
By doing that, we arrive at rigorous tight upper bounds
for quantum gates in the framework of linear optics quantum computation
with no feed-forward on the level of elementary gates.
In particular, it is proven that the non-linear sign shift gate can 
in fact be optimally realized with a success probability of exactly $1/4$.
Non-linear phase gates and equivalents in higher Fock layers
are also considered. These methods  will turn out to provide helpful 
tools, although the 
original problem has infinite dimension
 and is, to start with,  non-convex. 
The central difficulty here in the problem is that one
cannot bound the size of the auxiliary network a priori:
It may well be that large networks go in hand with a significant 
advantage \cite{DiffRemark}. 

Let us start by  stating the considered setting: we aim at
formulating a general recipe to find upper bounds for success probabilities
of gates of single modes preserving the energy using (i) photon sources, 
(ii) photon counters follows by postselection, and
(iii) static linear optical networks of any size, 
using an arbitrary number of auxiliary modes and photons and an arbitrary 
number of network elements, but without feed-forward on the level of 
individual gates (in which case 
the  unit probability as only upper bound is already 
known from the original 
\cite{KLM} and alternative schemes \cite{Cluster}). 
We will consider quantum gates  of the form
\begin{eqnarray}\label{Gate}
	|\psi_{\text{in}}\rangle = && y_0 |0\rangle + y_1 |1\rangle + 	...+ 
	y_N 
	|N\rangle\\
	\longmapsto && 
	U |\psi\rangle
	=
	(y_0   |0\rangle + y_1 e^{i\phi_1}
	|1\rangle + ...+ y_N  e^{i\phi_N}
	|N\rangle),\nonumber
\end{eqnarray}
where $|n\rangle$ denote the state vectors of number states and 
$\phi_1,...,\phi_N\in \RR$. To set the
phase $\phi_0=0$ merely corresponds to a change of the global phase and
does not restrict generality. 
This includes the important example of
the non-linear sign shift gate, acting as
\begin{eqnarray*}
	|\psi_{\text{in}}\rangle = && y_0 |0\rangle + y_1 |1\rangle +  y_2 
	|2\rangle 
	\longmapsto  
	(y_0 |0\rangle + y_1 |1\rangle - y_2 
	|2\rangle).
\end{eqnarray*}
In a static 
linear optical realization of the quantum gates, the gate can only 
be realized with a non-unity success probability.
Any network constisting of linear optical elements can
be decomposed into three steps, as has been
pointed out in Ref.\ \cite{ScheelLuetke} based on Ref.\ \cite{Reck} : (i) 
a preparation of a distinguished auxiliary  mode $2$ and all (unboundedly many) 
other auxiliary  modes jointly labeled $3$ in some initial pure state. 
(ii) A unitary operation 
of the input on $1$ and $2$, reflecting an application of a central
beam splitter 
with transmittivity $t\in[-1,1]$ 
(a convenient convention) and phase $\varphi\in [0, 2\pi)$. (iii) A measurement of all
modes labeled $2$ and $3$, associated with a  state 
vector $|\eta\rangle$.  
As a consequence, any optimal 
static linear optical network of a single input  
mode is reflected as a map 
%
\begin{eqnarray*}
	p_{\text{max}} U \rho_{\text{in}} U^\dagger=
	\langle\eta|
	(V_{1,2}\otimes \id_3) ( \rho_{\text{in}} \otimes 
	|\omega\rangle\langle \omega|)(V_{1,2}^\dagger\otimes \id_3)|\eta
	\rangle,
\end{eqnarray*}
for all input states 
$\rho_{\text{in}}=|\psi_{\text{in}}\rangle \langle 
\psi_{\text{in}}|$ of the input mode labeled with
$1$, $V_{1,2}$ is the unitary of the central beam 
splitter characterized by a real transmittivity $t$ and phase $\varphi$. 
Writing 
\begin{equation*}
	|\omega\rangle = \sum_{k=0}^{n} x_{k+1}  |k\rangle\otimes 
	|\omega_k\rangle
\end{equation*}
with real numbers $x_1,...,x_{n+1}$,
we have to require that 
\begin{eqnarray*}
	\sum_{k=0}^n x_{k+1}  f_k^{(j)} \varepsilon_{k+1}  = p^{1/2} e^{i\phi_j}
\end{eqnarray*}
for all $j=0,...,N$, 
with $\varepsilon_{k+1} = \langle \eta |k \rangle |\omega_k\rangle$ and 
\begin{eqnarray*}
	f_k^{(j)} = \langle j| \langle k | V_{1,2} |j\rangle |k\rangle = 
	e^{i\varphi j } g_k^{(j)},
\end{eqnarray*}
where the real $g_k^{(j)}$, introduced for convenience of notation, 
depend on $t\in[-1,1]$ only.

The problem is essentially now to find the optimal transmittivity 
$t\in[-1,1]$, phase $\varphi\in [0,2\pi)$,  
state vectors $|\eta\rangle$ and 
$ |\omega_0\rangle , ...,  |\omega_n\rangle$ for the 
optimal $n\in \NN$, and the optimal $x_1,...,x_{n+1}$ in 
order to bound the optimal success probability. This is as such a
very involved problem: The number $n$ 
cannot be bounded from above, meaning we cannot a priori bound the 
required resources 
in the network. This makes it formally an infinite-dimensional problem. 
The function we consider is not convex, so we may and are expected to 
encounter infinitely many local maxima. So even numerically, without truncating the 
problem cannot be solved as such. 
In order to circumvene these difficulties, 
two central ideas will be employed: We 
treat part of the objective variables as parameters in the problem, such that the 
remaining problem can be relaxed to a convex quadratic program. In this way
we can exploit methods from convex optimization. For
this resulting problem we make use of the ideas of Lagrange 
duality \cite{Semi} and outer approximations, 
and are able to explicitly construct a family of 
solution to the dual.
Let us first clearly state the strategy: 

(I) We consider the problem for each $t\in[-1,1]$, $\varphi\in [0,2\pi)$,
each $n\in\NN$, 
and all legitimate  $\varepsilon_1,...,\varepsilon_{n+1}$ as defined 
above. This choice will be denoted as 
$(t,\varphi,n, \varepsilon_1, ..., \varepsilon_{n+1})$.
(II)
We formulate the remaining 
problem of finding upper bounds to the 
optimal success probabilities as a quadratic optimization 
program, which can be relaxed to a semi-definite program   \cite{Semi}
in $x_1,...,x_{n+1}$. 
(III)  
Then, we are in the position to establish the dual problem. 
(IV) 
A family of explicit constructions of solutions of the dual will be
presented. (V) These solutions can be simplified such that
the dependence
on the specific choice of $\varepsilon_1,...,\varepsilon_{n+1}$ and $\varphi$
and $t$ can be 
eliminated. This will be done exploiting two key ideas: on the one hand, 
by using both families of solutions of the dual problem, dependent on 
$\varepsilon_1,...,\varepsilon_{n+1}$ and $\varphi$, on the other hand
by appropriate convex outer approximations. 
These powerful methods will allow us to identify 
rigorous general upper bounds for all numbers of 
auxiliary  modes, even though
the original problem is unbounded in size. In a sense, we approach the 
optimal solution 'from the other side'.

(I) The first simplification is that we may choose any 
$\varepsilon_1,...,\varepsilon_{n+1}$ for some $n$ 
satisfying 
$\sum_{k=1}^{n+1} (\alpha_k^2 + \beta_k^2)=1$, 
denoting with 
$\alpha_k,\beta_k\in \RR$ the real and imaginary parts of $\varepsilon_k$,
	$\varepsilon_k = \alpha_k + i \beta_k$. 
Again for simplicity of notation, we introduce
\begin{equation*}
	e^{i (j\varphi - \phi_j)} = \xi^{(j)}+ i 
	\zeta^{(j)}
\end{equation*}
with $\xi^{(j)}, \zeta^{(j)}\in \RR$. 
Success of the gate requires that
\begin{eqnarray}\nonumber
	\sum_{k=1}^{n+1} x_{k}  (\alpha_{k} \xi^{(j)}- \beta_{k} \zeta^{(j)}) 
	g_{k-1}^{(j)} &=&
	\sum_{k=1}^{n+1} x_{k}  (\alpha_{k} \xi^{(l)}- \beta_{k} \zeta^{(l)}) 
	g_{k-1}^{(l)},\\
	\sum_{k=1}^{n+1} x_k  (\beta_k \xi^{(j)} + \alpha_k \zeta^{(j)}) g_{k-1}^{(j)}
	&=& 0,\nonumber
\end{eqnarray}	
for $j,l=0,...,n$. Note that this is already a major simplification: instead of maximizing the actual
trace of the state, we set the imaginary part to zero and avoid a very involved
additional quadratic constraint at
this point, without losing generality.
The square of the quantities of the first line
in the previous equation is then the 
success probability.

(II) We have to optimize for all 
$(t,\varphi,n, \varepsilon_1, ..., \varepsilon_{n+1})$
over all weights $x_1,...,x_{n+1}$
satisfying 
	$\sum_{k=1}^{n+1} x_k^2=x^T x =1$.
This freedom corresponds to the weights in the preparation of the initial 
state of the auxiliary  modes. The fact that we cannot restrict the size of 
the linear optics network is here reflected by the fact that we have to 
optimize over all possible weights corresponding to different 
preparations, even over all $n$. 
In this form, however, we will see that the problem is 
handable. The constraint $x^T x=1$ 
can be relaxed to 
$x^T x\leq 1$, 
which is a convex quadratic 
constraint that can also be written as
\begin{equation*}
	\left[
	\begin{array}{cc}
	1 & x^T\\
	x & \id_{n+1,n+1}
	\end{array}
	\right]\geq 0.
\end{equation*}
So in general, the problem of assessing a bound for the optimal 
success probability can be reduced to the following {\it maximization 
problem} in the vector $x=(x_1,...,x_{n+1})$, reflecting the maximization 
of the success probability. The maximization problem in this vector 
(but not the full problem) is found to be 
manifestly of the form of a so-called 
semi-definite optimization problem \cite{Semi}. 
After a number of elementary steps, 
the maximization problem in this vector can be cast into
the following convenient form
of a maximization problem in the
real symmetric matrix $Z\in \RR^{(n+3)\times (n+3)}$  as
\begin{eqnarray}
\text{maximize} \,\, && -\tr[F_0 Z],\label{nowtheprob}\\
\text{subject to}\,\, && \tr[e_{a,a} Z]=1,\,\, a=1,...,n+3, \nonumber\\
	&& \tr[(e_{a,b}+ e_{b,a}) Z]=0,\,\,  a, b=3,...,n+3, \, a\neq b , \nonumber\\
	&& \tr[e_{1,a} Z]= \tr[e_{a,1} Z]=0, \,  a =2,...,n+3 , \nonumber\\
	&& \tr[F_j  Z]=0,\,\, 
	j = 1,..., 2N+2 .
	\nonumber\\
	&& Z\geq 0.\nonumber
\end{eqnarray}
The square of the solution is an upper bound for the success
probability. Here, $F_0 =\text{diag}(1,0,...,0)$, and
\begin{equation*}
	F_j= 0_{1,1} \oplus \left[
	\begin{array}{cc}
		  0 &  (c^{(j)})^T-(c^{(0)})^T\nonumber\\
	   c^{(j)} -  c^{(0)}  & 0_{n+1,n+1}\nonumber\\
	\end{array}	
	\right],\nonumber
\end{equation*}
$ j=1,...,N$,
correspond to the matrices that ensure the proper realization of the gate
on the level of the real part, and 
\begin{equation*}
	F_{j+N+1}=0_{1,1}\oplus 
	 \left[
	\begin{array}{cc}
	  0 &  (d^{(j)})^T\\
	  d^{(j)} & 0_{n+1,n+1}\\
	\end{array}	
	\right],
\end{equation*}
$ j=0,...,N$,
to the complex part, with $0_{k,l}$ denoting the 
$k\times l$ 
matrix all elements of which are zero.
 The matrix 
	\begin{equation*}
	F_{2 N+2} = \id_{1,1}\oplus 
	\left[
	\begin{array}{cc}
 	  0 &  -     (c^{(0)})^T/2  \\
	  -   c^{(0)}/2   & 0_{n+1,n+1}\\
	\end{array}	
	\right]
\end{equation*}
finally links the contraints in the primal 
problem. Here, the abbreviations
\begin{eqnarray*}
	c^{(j)} &=& ( (\alpha_0 \xi^{(j)} - \beta_0 \zeta^{(j)})
	g_0^{(j)} ,...,  (\alpha_n \xi^{(j)} - \beta_n \zeta^{(j)})
	g_n^{(j)}),  \\
	d^{(j)} & = & (   (\beta_0 \xi^{(j)} + \alpha_0 \zeta^{(j)})
	g_0^{(j)} ,..., (\beta_n \xi^{(j)} + \alpha_n \zeta^{(j)})
	g_n^{(j)}), 
\end{eqnarray*}
are used for $j=0,...,N$. The matrix $e_{a,b}\in \RR^{(n+3)\times 
(n+3)}$ denotes the matrix all 
entries of which are
zero, except of an entry $1$ at $(a,b)$.  The latter matrix
$	F_{2 N+2} $ can be replaced by
	\begin{equation*}
	G = \id_{1,1}\oplus 
	\left[
	\begin{array}{cc}
 	  0 &  -    \gamma  c^{(0)} /2  \\
	  -  \gamma  (c^{(0)})^T/2   & 0_{n+1,n+1}\\
	\end{array}	
	\right]
\end{equation*}
	with $\gamma\in [1,\infty)$ to be fixed later,
such that   $p_{\text{max}}$ corresponds to the square of
the optimal objective value of Eq.\ (\ref{nowtheprob}) 
for $\gamma=1$, and is smaller for
$\gamma>1$. This seemingly irrelevant modification will turn out to be
a  helful idea 
later on, to eliminate the dependence on the phase $\varphi$.
%

(III) We can now formulate the dual problem to this 
optimization problem delivering the bounds, as a solution can 
explicitly be constructed \cite{Standard}. It can be shown that the dual 
problem can be written as follows, which is now a {\it minimization 
problem } in the objective vectors $z\in \RR^{n+2}$,  $v\in \RR^{2N}$, and 
the matrix
$V\in \RR^{(n+3)\times (n+3)}$,
\begin{eqnarray*}
\text{minimize} \,\, && q^T z,\\
\text{subject to}\,\, && F_0 + \text{diag}(0,z_1,...,z_{n+2}) + 
\sum_{a=1}^{2N+1} 
	v_a F_a  \nonumber\\
	&+& 
	V + v_{2N+2}
	G
	\geq 0,\nonumber
\end{eqnarray*}
where $q=(1,...,1)$, and matrix $V$ has to be of the form 
$V=0_{2,2} \oplus W$,
with $W\in\RR^{(n+1)\times (n+1)}$ being a real symmetric matrix 
satisfying
$W_{a,a}=0$ for all $a=1,...,n+1$. In general, every
solution of a dual problem to a semi-definite problem gives
a bound to the optimal solution to the primal problem, 
as is not difficult to see \cite{Weak}. Once we are able to
construct a solution $z$ 
of the dual for all values of
$(t,\varphi,n, \varepsilon_1, ..., \varepsilon_{n+1})$, 
we arrive at a rigorous upper bound for the primal problem. 
As such, 
\begin{equation*}
 	p_{\text{max}}\leq (q^T z)^2  / \gamma^2
 \end{equation*}
gives an upper general bound of the desired success probability.

(IV) We will now explicitly construct a family of solutions, dependent on 
a single number $\delta \in \RR$.
The presented solutions may look like unlikely objects, yet, 
they will deliver the desired bounds. To construct the family of solutions 
for the dual amounts to 
finding appropriate values for a matrix of arbitrary dimension. The 
intuition behind the construction draws from two essential observations:
on the one hand, the problems in $x$ and $(\varepsilon_1,...,\varepsilon_n)$
are two intertwined quadratic problems. So the solutions can be constructed 
such that the dependence from $(\varepsilon_1,...,\varepsilon_n)$ is 
entirely cancelled. The other central idea, dealing with the very involved 
constraints provided 
by polynomials of arbitrary order in $t$, is to get bounds by 
appropriately constructing provable outer approximations. This structure of 
the problem we encounter here is not only specific for the optimization problem
at hand, but expected to be a generic feature in problems related to 
linear optics: roughly speaking, 
the intertwined quadratic problems originate from the auxiliary
systems, whereas the polynomial constraints of high order from the
distinguished passive optical element.

In the construction, to start with, we choose 
$v_{2N+2}=1$. 
Let for convenience 
$w\in \RR^{n+1}$ be defined as
\begin{eqnarray*} 
	w&=& \bigl(-\gamma /2 - \sum_{j=1}^{N} v_j \bigr)  c^{(0)} 
	+ \sum_{j=1}^N  v_j  c^{(j)} 
	+ \sum_{j=0}^N v_{N+j+1}  d^{(j)},
\end{eqnarray*}
We are free to choose 
\begin{eqnarray*}
	v_{j} &=& - \cos(j \varphi) s_j, \\
	v_{N+j+1}& = &\sin(j \varphi)  s_j, 
\end{eqnarray*}
$j=1,...,N$, with  functions $s_j:[-1,1]\rightarrow \RR^+ $ yet to be specified.
This freedom will later give rise to the outer approximation.
Then, let us set
\begin{equation*}
	\gamma = 2 \sum_{j=1}^N s_j (1-\cos(\varphi j)) +1.
\end{equation*}
This means that always  $\gamma\geq 1$, which is used to eliminate 
the unwanted dependence
of $\varphi$.  That is,
\begin{eqnarray*}
 	w_{k}/\alpha_k =    (-1 /2 + \sum_{j=1}^N  s_j  )
 	g_k^{(0)} -  \sum_{j=1}^N \cos(\phi_j) s_j g_k^{(j)}  ,
\end{eqnarray*}
The matrix $W\in\RR^{(n+1)\times (n+1)}$  is taken to 
be of the form 
\begin{equation*}
	W_{a,b}=\left\{
	\begin{array}{ll}
	 w_a w_{b} ,\, &\text{ if } b\neq a,\\
	0,\, &\text{ if } b=a.
	\end{array}\right.
\end{equation*}
This construction yields a positive $V$ \cite{Proofsketch}.
%
Finally, we choose
\begin{equation*}
	z_a= \alpha_{a-1}^2 \delta
%
\end{equation*}	
for $a=2,..., n+2$ and $z_1= \delta$. 
	With this choice,  indeed  
\begin{eqnarray*}	
	F_0 + \text{diag}(0,z_1,...,z_{n+2}) + \sum_{a=1}^{2N} 
	v_a F_a +
	V +  v_{2N+2 }
	G \geq 0, 
\end{eqnarray*}	
	holds, so it is in fact a solution of the dual \cite{Proofsketch,Weak}.
This choice will indeed turn out to give
the appropriate upper bounds. 

(V) If we can now find functions $s_1,...,s_N:[-1,1]\rightarrow \RR^+$ 
such that there is a $\delta \in[0,1]$ with  
\begin{equation}\label{thaq}
	|w_k/\alpha_k| \leq \delta
\end{equation}
for all $k=0,...,n$, we can in fact eliminate the dependence on
$\alpha_{1},...,\alpha_{n+1}$ and $t$, as we have 
then an outer approximation of the feasible set. The outer
approximation defined by Eq.\ (\ref{thaq}) takes care of the
polynomial constraints in $t\in[0,1]$ of arbitrary order, 
constraints of a type that one would encounter in any 
optimization involving passive optical elements.

We have then indeed established an
upper bound: The above constructed solution yields
\begin{equation*}
	p_{\text{max}}\leq  (q^T z )^2/\gamma^2  \leq  
	(q^T z )^2\leq (
	\delta
	+ \sum_{i=1}^{n+1} \alpha_i^2 \delta)^2 \leq 4 
	\delta^2.
\end{equation*}
so $ p_{\text{max}}\leq 4 \delta^2$ 
is a rigorous upper bound for the success probability. So finding an
upper bound for the success probability amounts to finding solutions, 
possibly dependent on $t\in[-1,1]$, for 
$s_1,...,s_N$ such that Eq.\ (\ref{thaq}) is satisfied. 
This provides a general method that can be applied to all of the above 
considered gates. It is important to note that although we had the freedom
to construct this particular solution without caring whether this solution is
unique or even optimal, 	this implies a rigorous bound for the primal problem, 
and therefore for the optimal success probability. This gives rise to a 
recipe for finding upper bounds for success probabilities for all the above quantum gates using 
linear optics.

The example of the non-linear sign shift is 
on the one hand instructive to exemplify the general strategy, 
and on the other hand already the  
practically most important case. Here we have that $N=2$ and 
	$\phi_0=1$,
	$\phi_1=1$, and
	$\phi_2= \pi$. 
For this case of $N=2$, one finds
	$g^{(0)}_k = t^k$, 
	$g^{(1)}_k = 
t^{k-1}(t^2 - k (1-t^2))$, and 
	$g^{(2)}_k  = 
t^{k-2} (
	t^4 - 2k t^2 \, (1-t^2) +
	(1-t^2)^2 k(k-1)/2 )$ using standard expressions for 
the unitaries of beam splitters in the number state basis.
We now have to show that for each $t\in[-1,1]$
we can find $s_1,s_2:[-1,1]\rightarrow \RR^+$ such that Eq.\ (\ref{thaq})
is satisfied. More specifically, for all $t\in[-1,1]$
we have to have find $s_1,s_2$ such that
\begin{equation*}
-1/4 \leq (-1/2 + s_1+s_2) g^{(0)}_k 
- s_1 g_k^{(1)} + s_2 g_k^{(2)}\leq 1/4
\end{equation*}
for all $k=0,...,\infty$, so we have that $b=1/4$. 
Such a choice is given by 
\begin{equation}\label{FunnyChoice}
	(s_1,s_2)=\frac{1}{4}\left\{
	\begin{array}{ll}
	(1/(1-t),0)  , & \text{ if }t\in[-1 ,1-\sqrt{2} ),\\
	(0, 1/(1+t^2) ) , & \text{ if }t\in[1-\sqrt{2} ,0),\\
	(1,1/2) 	, & \text{ if }t\in[0 ,1),\\
	\end{array}\right.
\end{equation}
%
%
%
for all $k=0,...,\infty$ \cite{Actual}. 
This can be shown with elementary methods, on the
basis of only the functions in Eq.\ (\ref{FunnyChoice})
such that Eq.\ (\ref{thaq}) holds for 
$\delta=1/4$ for all $k$. 
This finally demonstrates
that the optimal success probability of a linear optical implementation of 
the non-linear sign shift gate without feed-forward is indeed $1/4$:
there are known schemes that fulfill this bound. This settles the question of 
the optimal success probability of this key quantum gate
in this setting.
This statement is interestingly completely independent of the network size, as long as 
it includes at least two auxiliary modes. The surprising result is that more resources
do not help at all, and the smallest known functioning scheme can already be proven to be
optimal. This unexpected outcome may also be taken
as a further motivation to 
further investigate hybrid solutions, 
slightly leaving the setting of linear optics \cite{Hybrid}.

The presented method can immediately be applied to
assess optimal success probabilities of other quantum gates
within the paradigm of linear optics. In order to 
exemplify the versatility of the approach, let us
finally investigate two further quantum gates: this 
is on the one hand the non-linear phase shift
gate, acting as 
\begin{equation*}
y_0 |0\rangle + y_1 |1\rangle
+ y_2 |2\rangle \longmapsto y_0 |0\rangle + y_1 |1\rangle
+ e^{i \phi_2} y_2 |2\rangle
\end{equation*}
 with some phase $\phi_2\in[0,2\pi)$. Here, the
presented method delivers immediately 
\begin{equation*}
p_{\text{max}}\leq (3 - \cos(\pi - \phi_2))^2/16,
\end{equation*}
consistent with $p_{\text{max}}=1$ for $\phi_2=0$ 
and $p_{\text{max}}=1/4$ for $\phi_2=\pi$: it hence
depends on the phase how difficult it is to
implement the gate. On the other 
hand, for the three photon gate 
\begin{equation*}
y_0 |0\rangle + y_1 |1\rangle + y_2 |2\rangle 
+ y_3 |3\rangle \longmapsto  y_0 |0\rangle + y_1 |1\rangle
+   y_2 |2\rangle - y_3 |3\rangle
\end{equation*}
we find that $p_{\text{max}}
\leq
1/9$. This indicates that for higher Fock layers, the optimal
success probabilies becomes even smaller. The implications
for a number of further gates including the CNOT will be presented in a forthcoming
publication. The key point is that this method allows one 
to argue without having to restrict the amount of allowed resources or the size of 
the specific network realizing a scheme. Moreover, a finite 
number of rounds of  
measurements and feed-forward can in principle be incorporated in such a setting.
Statements on the distinguishability using auxiliary systems \cite{Loock}
are also accessible. As such, these ideas are hoped to be
useful to contribute to finding linear optical 
schemes that make use of the minimal resources, and  to
bringing linear optics
quantum computation closer to feasibility. 

Discussions with W.J.\ Munro, 
S.\ Scheel, P.\ van Loock, C.\ Emery, 
J.D.\ Franson,
K.\ Nemoto, 
P.\ Kok, M.A.\ Nielsen,  
N.\ L{\"u}tkenhaus, and M.\ Knill
are warmly acknowledged. This work was supported by
the DFG
(SPP 1078), the EU (IST-2001-38877), and 
the EURYI Award Scheme.



\begin{thebibliography}{99}
\bibitem{Standard}
	The most general form of a 
	semi-definite program is the maximization of 
	$-\text{tr}[Z F_0]$ subject to $\text{tr}[Z F_i]=c_i$ 
	for $i=1,...,n$ and $Z\geq 0$. In turn, its dual problem is
	the minimization of 
	$c^T x$ subject to $F_0 + \sum_{i=1}^n x_i F_i\geq 0$.
\bibitem{Weak}
	This is the content of so-called weak duality. In the notation of
	footnote \cite{Standard}, this can be seen as follows: let 
	the matrix $Z\geq 0$ be the optimal solution of the primal problem,
	and the vector $x$ be any solution of the dual. Then, 
	$-\text{tr}[Z F_0] \leq \sum_{i=1}^n	
	x_i \text{tr} [Z F_i] = c^T x$, delivering 
	an upper bound.
\bibitem{Proofsketch}
	This follows from the fact that
	$X\geq 0$ for a matrix $X$ all entries of which are
	$1$, and that therefore
	$\text{diag} (r_1,...,r_{n+1}) X   \text{diag} (r_1,...,r_{n+1})\geq 0$
	for any $r_1,...,r_{n+1}\in \RR$.
\bibitem{Actual}
	It is known that with a  choice of $t=1-\sqrt{2}$ for the transmittivity
	of the  central beam splitter a scheme realizing $1/4$ success probability 
	can be constructed.
	Interestingly, there are signatures of this optimal choice in the solution for the
	dual problem.
\end{thebibliography}
\end{document}